\documentclass[11pt]{article}

\newcommand{\bmx}{\mbox{\boldmath $x$}}
\newcommand{\bbeta}{\mbox{\boldmath $\beta$}}

\newcommand{\bmy}{\mbox{\boldmath $y$}}

\large\normalsize
\usepackage{indentfirst}
\usepackage{amssymb}
\usepackage{graphicx}
\usepackage{multirow}
\usepackage{array, rotate}
\usepackage{figsize,subfigure,ifthen,calc,psfrag,epsfig,graphicx}
\usepackage{enumerate,amsthm,amsmath,amssymb,amsxtra,comment}
\usepackage{booktabs, hyperref}
\usepackage[round]{natbib}
\usepackage{psfrag}
\usepackage{epsfig, appendix}

\begin{document}

\title{\bf Bayesian Mode Regression}
\author{Keming Yu\thanks{{\it Address for correspondence:} Department of Mathematical Sciences, Brunel
University, UK. \newline E-mail: keming.yu@brunel.ac.uk
}\\ {\em\small  Brunel University, UK  }\\ Katerina Aristodemou
\\
{\em\small Brunel University, UK}}
\date{}
\maketitle


\noindent {\bf Summary.} \,
Like mean, quantile and variance, mode is also an important measure of central tendency and data summary. Many practical questions often focus on ``Which element (gene or file or signal) occurs most often or is the most typical among all elements  in a network?". In such cases mode regression provides a convenient summary
of how the regressors affect the conditional mode and is totally different from other regression models based on conditional mean or conditional quantile or  conditional variance.
   Some inference methods have been used for mode regression but none of them from the Bayesian perspective. This paper introduces  Bayesian mode regression by exploring three different approaches. We start from a parametric Bayesian model by employing a likelihood function that is based on a mode uniform  distribution. It is shown that irrespective of the original distribution of the data, the use of  this special uniform distribution is a very natural and effective way for Bayesian mode regression. Posterior estimates based on this parametric likelihood, even under misspecification, are  consistent and asymptotically normal. We then develop  a nonparametric Bayesian model by using Dirichlet process (DP) mixtures of mode uniform distributions and finally we explore Bayesian empirical likelihood mode regression by taking empirical likelihood into a Bayesian framework. The paper  also demonstrates that a variety of improper priors  for the unknown model parameters yield a proper
joint posterior. The proposed approach is illustrated using simulated datasets and a real data set.

\noindent{\bf Keywords:} Bayesian inference; empirical likelihood; Markov Chain Monte Carlo
methods; mode regression; nonparametric Bayesian; parametric Bayesian.

\newpage

\section{Introduction}

Mode, the most likely value of a distribution, has wide applications in  biology, astronomy, economics and
finance. For example, it is not uncommon in many fields to encounter data distributions that are skewed or
contain outliers. In those cases, the arithmetic mean may not be an appropriate statistic to represent the
center of location of the data. Alternative statistics with less bias are the median and the mode.
The mean or median of two densities may
be identical, while the shapes of the two densities are quite different. Mode preserves some of the
important features, such as wiggles, of the underlying distribution function, whereas the mean or median tend to
average out the data. Actually, as an important statistic, mode has been used in modern science to identify the most frequent or the most typical element in certain network systems  (\citet{hedges2003}, \citet*{heckman2001}, \citet*{kumar1998}, \citet{Markov1997}).

A mode estimator is often defined as the maximum of the estimated distribution density, typically under
nonparametric kernel estimation. Such mode estimation has attracted a lot of attention in the statistics literature for decades  by various authors
(\citet{yasukawa1926}, \citet{parzen1962}, \citet{grenander1965},  \citet{eddy1980},   \citet{bickel1996}, \citet{birge1997},
\citet{berlinet1998} and \citet{meyer2001} among others).  Similarly,  conditional mode estimation  is typically carried out by conditional density estimation  via different nonparametric methods (\citet{gasser1998}, \citet{hall2001a} and  \citet{hall2001b}, \citet{brunner1992}, Ho (2006), \citet{dunson2007} among others).
 However, these nonparametric conditional density based mode regression models are not a direct estimation of the conditional  mode. The problem with these methods is twofold: the
estimation of the conditional density may suffer from the well-known ``curse of
dimensionality'' and,  it is  hard to describe and interpret the estimated conditional mode in terms of predictors or covariates.
 Direct inference for mode regression was  explored by \citeauthor*{lee1989} first in 1989 and then
 in 1993 (\citeauthor {lee1989},\citeyear{lee1989},\citeyear{lee1993}). However, it has not been well-applied due to lack of proper inference tools.
Recently, \citeauthor*{kemp2012}(2012)
 relaxed  Lee's restriction on truncated dependent variables  and employed alternative   kernel estimation. However, their regression coefficient estimation
 has slow convergence rate, involves bandwidth selection and provides only approximate normal confidence intervals. Moreover,    direct Bayesian method for mode regression is not available but there is clear practical
motivation from this perspective.

In this paper we introduce a fully Bayesian framework for direct mode regression inference by using three approaches: a parametric Bayesian method, a nonparametric Bayesian method and an empirical likelihood based Bayesian method.
The remainder of the paper is organized as follows. Section \ref{section: BMR} introduces the three approaches, describes the theoretical and computational framework of these methods and  gives their mathematical justification. In Section \ref{section:Applications} we illustrate the proposed methods through two simulated case-studies and a real example.  We conclude with a short discussion in Section \ref{sec: Conclusions}.

\section{Bayesian mode regression}
\label{section: BMR}

Consider an arbitrary random variable $Z$, let $F_Z(z)$ be the distribution of $Z$ with density $f_Z(z)$ and let $K(Z;\cdot)$ be the {\it step-loss function} \citep{Manski1991} such as,
\begin{equation}
K(Z; \mu)= I[\frac{|Z-\mu|}{\sigma}\geq 1],
\end{equation}
with $\sigma>0$ and $I[A]$ being the indicator function of event $A$.
If $f_Z(z)$ is
symmetric about $\mu$ or if $\mu$ is the middle value of the interval of length $2\sigma$ that captures the most probability under $F_Z(z)$ then $\hat\mu=\mbox{argmin}_\mu E\{L(Z; \mu)\}$ is the mode of $Z$.
Lee (1989) introduced mode regression, or the conditional mode of $y$ given $\bmx$, as $mode(y|\bmx)=\bmx' \bbeta$ based on the loss function
$K(y; \bmx' \bbeta),$ where $\bbeta$ is the regression parameter.  That is, given  a sample $\{(\bmx_1, y_1), \,...\,(\bmx_n, y_n)\}$ from $(\bmx, y)$, when $\sigma$ approaches $0$,
the  parameter $\bbeta$ in the conditional model of $y|\bmx$ is estimated by
\begin{equation}
\label{eq: Loss function}
\hat{\bbeta}=\mbox{argmin}_{\bbeta} \sum_{i=1}^nK(y_i;\, \bmx_i'\bbeta)
\end{equation}

\subsection{Parametric Bayesian  method}
\label{subsection: Bayesian setting}

The conditional mode denoted as $mode(y|\bmx)=\bmx' \bbeta$  can  be  reformulated as a standard regression model
\begin{equation}
\label{eq: linear model}
y=\bmx'\bbeta+\epsilon
\end{equation}
with zero mode for the error term $\epsilon$.

 Given  a sample $\{(\bmx_1, y_1), \,...\,(\bmx_n, y_n)\}$ from $(\bmx, y)$, note that
$\hat{\bbeta}=\mbox{argmax}_{\bbeta}\sum_{i=1}^n I[|y_i-\bmx_i' \bbeta|\leq\sigma].$
That is, $\hat{\bbeta}$ in (\ref{eq: Loss function}) can be regarded as
the maximum likelihood estimates of the ``working'' likelihood function
\begin{equation}
\label{equation: likelihood}
L(y | \bbeta) \propto \sigma^{-n}\,\sum_{i=1}^n I(|y_i-\bmx_i' \bbeta|\leq \sigma).
\end{equation}
Therefore, the Bayesian mode regression estimates, denoted as $\hat \bbeta_B$, can be obtained using the posterior distribution of $\bbeta$,
\begin{equation}
\label{equation: Posterior}
\pi(\bbeta | y) \propto L(y | \bbeta) \, \pi(\bbeta),
\end{equation}
where $\pi(\bbeta)$ is the prior distribution of $\bbeta.$
Although a standard conjugate prior distribution
is not available for the mode regression formulation, MCMC methods may be used for extracting the posterior
distributions of $\bbeta$.

\subsection{Consistency and asymptotic normality}
\label{subsec: Consistency}

The classical mode regression parameter estimator $\hat{\bbeta}$ of \citeauthor {lee1989} (\citeyear{lee1989},\citeyear{lee1993})
\[\hat{\bbeta}= argmax_{\bbeta}\frac{1}{n} \sum_{i=1}^n I[|y_i-\bmx_i' \bbeta|\leq\sigma]. \]
is known to be consistent. According to \citet{white1982}, any posterior estimator $\hat \bbeta_B$ from the likelihood function (4) with a flat prior, even if misspecified,  is still consistent, in the sense of minimization of   the
Kullback-Leibler distance between the true distribution and the parametric family to which
the approximation belongs or in the spirit of the quasi-maximum likelihood estimator of \citet{white1982}.

Further, although Bayesian inference does not require large sample theory we provide evidence that the posterior distribution obtained via the proposed Bayesian approach, under certain regularity conditions,  is asymptotically normal when the sample size increases. This  is the same as  the classical  mode regression estimator $\hat{\bbeta}$ whose  asymptotic normality was derived under  a special case of ``M-estimators'' (\citet{huber1973},  \cite{lee1993}).

  In fact,
let $I(\bbeta)=E[\{\frac{\partial}{\partial \bbeta}\log f(y|\bbeta)\}^2]$ be the total Fisher information in the data and define
$I_n(\hat\bbeta_B)=I(\bbeta)_{\bbeta=\hat\bbeta_B}$. Under certain regularity conditions, Taylor power series expansions of the logarithm of the posterior distribution leads to \[\log \pi(\bbeta|y)= \log \pi(\hat\bbeta_B|y)-\frac{1}{2}(\bbeta-\hat\bbeta_B)^T
I_n(\hat\bbeta_B)(\bbeta-\hat\bbeta_B),\]
hence, \[\pi(\bbeta|y)\approx\pi(\hat\bbeta_B|y)\,\exp[-\frac{1}{2}(\bbeta-\hat\bbeta_B)^T
I_n(\hat\bbeta_B)(\bbeta-\hat\bbeta_B)]\]
\[\propto\exp[-\frac{1}{2}(\bbeta-\hat\bbeta_B)^T I_n(\hat\bbeta_B)(\bbeta-\hat\bbeta_B)],\]
which is the kernel of a $N_p(\bbeta| \hat\bbeta_B,\,I_n^{-1}(\hat\bbeta_B))$ density. This  implies that  $\sqrt{n}(\hat\bbeta_B-\bbeta)\sim N(0,\,I_n^{-1}(\hat\bbeta_B))$, where $I_n^{-1}(\hat\bbeta_B)$ is specified  below, which has the same form as the asymptotic distribution of ``M-estimators". On the other hand, the likelihood function associated with Bayesian mode regression can also be formulated as
\[L(\bbeta)=\frac{e^n}{(2\sigma)^n}\,\prod_{i=1}^n\exp(- I[|y_i-x_i'\bbeta|\leq\sigma]),\]
which is a likelihood function based on the uniform probability density
\begin{equation}
\label{equation: Mode density}
f_{\sigma}(u)=\frac{e}{2 \sigma} \, \exp(-I[|u-\mu|\leq\sigma]),
\end{equation}
for a window parameter $\sigma>0$.
Then, under a flat prior, from
\[f(y|\bmx, \bbeta,\sigma)=\frac{e}{2\sigma}\exp[-I[|y-\bmx'\bbeta|\leq \sigma]],\]
we have
\[\log f(y|\bmx, \bbeta, \sigma)=1-\log(2\sigma) -I[|y-\bmx^T\bbeta|\leq \sigma],\]
\[\frac{\partial}{\partial \bbeta}\log f(y|x, \bbeta, \sigma)=-I[|y-\bmx^T\bbeta|\leq \sigma],\]
\[I_n(\hat\bbeta_B)=E[\{\frac{\partial}{\partial \bbeta}\log f(y|\bbeta)\}^2]|_{\hat\bbeta_B}=E\{I[|y-\bmx^T\bbeta|\leq \sigma]\}_{\hat\bbeta_B}.\]
Thus, the asymptotic justification of using the proposed ``working'' likelihood for parametric Bayesian mode regression is fully outlined.

\subsection{The estimation of covariance matrix of classical estimates }

Under the classical approaches of \citeauthor{lee1989} (\citeyear{lee1989}, \citeyear{lee1993}) and \citet{kemp2012}, the covariance matrix, $cov\{\hat{\bbeta}\}$ of the classical estimator $\hat{\bbeta}$ and its inverse are often required but difficult to estimate or compute numerically, especially under small or moderate samples. A by-product of the proposed Bayesian approach is that using the MCMC posterior sample leads to a natural and  efficient estimation of $cov\{\hat{\bbeta}\}$ and other asymptotic quantities of $\hat{\bbeta}$.

In fact, a MCMC scheme constructs a Markov chain whose equilibrium distribution is the joint posterior, $p(\bbeta|data)$. After running the Markov chain for a burn-in period, one obtains samples from the limiting distribution, provided that the Markov chain has converged. Given that the chain has converged, the frequency of appearance of the parameters in the Markov chain represents their posterior distribution. An informative full density distribution of the model parameters is readily obtained rather than a single point estimate as in the classical approach.

When a Markov chain, \textit{S},  is drawn from the posterior
distribution, $p(\bbeta|data)$: $S=(\bbeta^{(1)},\bbeta^{(2)},...,\bbeta^{(N)})$, where N is the number of
draws after burn-in, a consistent estimate of the inverse of the covariance matrix $cov\{\hat{\bbeta}\}$  can be obtained by multiplying by $N$ the variance-covariance matrix of this MCMC sequence \citep{chernozhukov2003}.

\subsection{Prior selection  and proper posteriors}

In this  section we  demonstrate that in the absence of any realistic information one could use improper uniform prior distributions for all the components of $\bbeta$ as such a choice yields a proper
joint posterior and then we address the important issue of specifying a prior for parameter $\sigma$.

Below, we show that if we choose the  prior of $\bbeta$ to be improper uniform, then the resulting
joint posterior distribution will be proper.

\noindent {\bf Theorem 1:} If the likelihood function is given by
(\ref{equation: likelihood}) and $\pi(\bbeta) \propto 1$, then the posterior distribution of $\bbeta$, $\pi(\bbeta | \bmy)$,
will be a proper distribution. In other words
\[
0 < \int\,\pi(\bbeta|\bmy)\,d\bbeta < \infty,
\]
or, equivalently,
\[
0 < \int\,L(\bmy|\bbeta)\,\pi(\bbeta)\,d\bbeta < \infty.
\]

The proof can be found in \ref{appendix}.

In practice one usually assumes that the components of $\bbeta$ have independent improper uniform prior
distributions which is a special case of the above theorem.

Next we address the issue of determining a suitable prior for the parameter $\sigma$.  The aim is to manage to
determine a value that is neither too small nor too large to prevent underutilization of data with
\(|y_i-x_i' b|<\sigma\) or with \(|y_i-x_i' b|>\sigma\) respectively.  \citeauthor*{lee1989} (\citeyear{lee1989}, \citeyear{lee1993}) suggested  some  possible methods for determining the value of $\sigma$ (including trial and error and bootstrapping methods).

In this work we apply a Uniform$(w_1,w_2)$ prior on $\sigma$, where $w_i$ can be determined using one or more of the following three rules-of-thumb, depending on the assumption for the underlying distribution.

\begin{itemize}
 \item The empirical rule, which states that, given a symmetric distribution, approximately 99.7\% of
     the data values fall within three standard deviations (sd) of the mean, therefore,
     $w_i=3*\widehat{sd}$;
\item	Chebyshev's Theorem, which is true for any sample set no matter what the distribution is,
and states that at least 93.75\% of the data values fall within four standard deviations of the mean,
therefore, $w_i=4*\widehat{sd}$;	
\item	Variations of Silverman's plug-in estimate for the bandwidth \citep{silverman1986}, a simple formula for $w_i$ that depends on the sample size $n$ and the sample standard deviation $sd$, given by $w_i=1.3643\delta n^{-0.2}[min(\widehat{sd},IQR/1.349)]$ where $IQR$ is the sample inter quantile range and $\delta=1.3510$ for a uniform kernel. This formula assumes data which is normally distributed and uses $IQR/1.349$ as an alternative estimate of $w_i$ that protects against outliers. These plug-in estimates for $w_i$ work well in practice, especially for symmetric unimodal densities even if the data is not normally distributed. Alternatively, $IQR/1.349$ can be replaced by $1.4826*MAD$ to cover data with large number of outliers.
 \end{itemize}

However, the choice of a suitable prior for $\sigma$ can be difficult in practice, therefore with the aim of developing  a more flexible model, in the following section, we relax the distributional assumption on the prior for $\sigma$ using a Dirichlet process prior. This leads to a flexible nonparametric mixture model. The method is nonparametric in the sense that we do not assume that the prior belongs to any fixed class of distributions.

\subsection{Nonparametric Bayesian method}
\label{sec: Semiparametric approach}

In this section, we formulate a nonparametric Bayesian mode regression model to avoid critical dependence on the mode uniform  distribution assumption thus to address the issue of misspecification that may arise under the parametric Bayesian method.

A density $f(\cdot)$ on  $\mathbb{R}^{+}$ is non-increasing if and only if there exists a distribution function $G$ such that $f(x|G)=\int \sigma^{-1}I_{[0<x<\sigma]}dG(\sigma)$ \citep{feller1971}. Therefore, any unknown density $f(\cdot)$ (with mode $\theta$), symmetric or not, can be represented as a scale mixture of symmetric uniform distributions, that is

\begin{equation}
\label{equation:asymmetric density}
f(x|\theta, G)=\int \frac{1}{2\sigma}I_{[-\sigma<x-\theta<\sigma]}dG(\sigma),
\end{equation}
where $G$ is the mixing distribution supported on $\mathbb{R}^{+}$.

Then a nonparametric Bayesian mode regression model can be expressed in the hierarchical form
\begin{equation}
\label{semiparametric}
\begin{array}{l}
y_i |\bbeta, \sigma_i \mathop{\sim}\limits^{ind} f(y_i-x_i'\bbeta;\sigma_i),i=1 \cdots n\\
\sigma _i |G \mathop{\sim}\limits^{iid} G, i=1 \cdots n \\
G|M,d\sim DP(M,G_{0}(\cdot,d))\\
\bbeta, M,d\sim p(\bbeta),p(M),p(d),
\end{array}
\end{equation}
where, $G$ is the mixing distribution, with base distribution $G_{0}$ and  concentration parameter $M$ and $f(y_i-x_i'\bbeta;\sigma_i) = \frac{1}{2\sigma} I_{[-\sigma<y_i-x_i'\bbeta<\sigma]}$ is the density of a uniform distribution on $(-\sigma,\sigma)$.

We take a uniform distribution as the base distribution, $G_0$, uniform prior for $M$ and we choose non-informative Normal priors for all the components of $\bbeta$.

\subsection{Empirical likelihood based Bayesian method}
\label{Bayesian EL for mode}

In addition to parametric and nonparametric likelihood, an empirical likelihood based method could be an alternative for Bayesian mode regression.  To derive an empirical likelihood for mode regression we begin with notations and a moment restriction. \citet{lee1993} generalized the mode regression estimator of  \citet{lee1989},  $\hat \bbeta=argmin_{\bbeta} E\{L(Y-\bmx' \bbeta)\},$  by using the rectangular kernel
$L(Y; \mu)=\{(\sigma^2-(Y-\mu)^2)I[|Y-\mu|<\sigma]\}.$
 Therefore, the moment restriction for the empirical likelihood can be obtained by  the derivative $\frac{\partial}{\partial \mu} L(Y; \mu)=2 (Y-\mu)I[|Y-\mu|<\sigma]$. Let $l(\mu)$ be the `derivative' of $L(.;\mu)$, then the mode, $\mu$, of $Y$ satisfies the moment restriction
$E(l(\mu))=0$, where $l(\mu)=(Y-\mu)\,I(|Y-\mu|<\sigma)$.

Under an empirical likelihood for mode regression $\mu=\bmx'\bbeta$, thus for any proposed $\bbeta$ to estimate the true $p$ dimensional  $\bbeta_0$ via empirical likelihood, we use the vector  estimating functions $g(X, Y, \bbeta)$ with component $g_j(X, Y, \bbeta)=l(X, Y, \bbeta)\,X_j$ for $j=1,.., p.$  Then, the
profile empirical likelihood ratio is  given by
\[\mathfrak{R}(\bbeta)=max\{\prod_{i=1}^n (n\,p_i)| \sum_{i=1}^n p_i\,g(X_i, Y_i, \bbeta)=0,\,p_i\geq 0,\,\sum_{i=1}^n p_i=1\}.\]
By a standard Lagrange multiplier argument we have

\[\mathfrak{R}(\bbeta)=\prod_{i=1}^n \{n\,p_i(\bbeta)\},\]
with the weights $p_i(\bbeta)=\frac{1}{n(1+\lambda' g(X_i, Y_i, \bbeta))},$ where the Lagrange multiplier $\lambda$ satisfies \[\sum_{i=1}^n \frac{g(X_i, Y_i, \bbeta)}{1+\lambda^{T}\,g(X_i, Y_i, \bbeta)\,}=0.\]

According to \citet{qin1994}, among others, the existence and uniqueness of $\lambda$  are guaranteed when the following
two conditions are satisfied: (1) zero belongs the convex hull of $\{g(X_i, Y_i, \bbeta)$, $i=1, ..., n\}$ and (2) the matrix
$\sum_{i=1}^n \{g(X_i, Y_i, \bbeta) g(X_i, Y_i, \bbeta)'\}$ is positive definite.

Under Bayesian inference we consider the empirical likelihood function $\mathfrak{R}(\bbeta)/n^n=\prod_{i=1}^n \{p_i(\bbeta)\}$, which can be combined with a prior specification $\pi(\bbeta)$ on the parameter $\bbeta$ to obtain the posterior distribution
\[\pi(\bbeta|data) \propto \pi(\bbeta)\,\mathfrak{R}(\bbeta).\]

\subsection{Asymptotic properties of Bayesian empirical likelihood}

Before studying the asymptotic normality of the empirical likelihood based Bayesian mode regression parameter estimates, we should provide the consistency of the empirical likelihood estimator, which is a necessary condition for the
asymptotic normality of the posterior.  As the criterion function $g(X,Y,\bbeta)$ results in  non-smooth estimating equations we employ a similar method to the one use by \citet{molanes2009}, among others, to derive our asymptotic results.

Let $\hat{\bbeta}= argmax_{\bbeta}  \mathfrak{R}(\bbeta)$ be the empirical likelihood estimates in a compact set of parameter space which contains the true parameter $\bbeta_0$. Then note that our criterion function $g(X,Y,\bbeta)$ can be regarded as a special case of M-estimators as discussed in Chapter 5 of \citet{van1998} and satisfies the conditions of theorem 5.7 in the book. Under some regular conditions such as uniformly continuous and bounded imposed on  the marginal distribution of $X$ and conditional distribution of $Y$ given $X$,  and  assume  the matrix
 $E\{g(X, Y, \bbeta)\,g(X, Y, \bbeta)'\}>0$, then  $E\{g(X,Y,\bbeta)\}$ is sufficiently smooth  in a compact set of parameter space which contains  $\bbeta_0$, so the consistency condition $C_3$ of \citet{molanes2009} holds, that is,  the consistency of  $\bbeta's$ empirical likelihood estimates is established.

The asymptotic normality of the posterior distribution $\pi(\bbeta|data)$ could be established using  the fact that the empirical log-likelihood ratio for $\bbeta$ is  well approximated by certain quadratics in the sense of Lemma 6 of \citet{molanes2009} so that,
\[\Gamma_n(\bbeta) \equiv -n^{-1}\sum_{i=1}^n \log(1+\lambda^{T} g(X_i, Y_i, \bbeta))\]
\[=-\frac{1}{2}(\bbeta-\bbeta_0)' V_{12}V{11}^{-1}V_{12}(\bbeta-\bbeta_0)+n^{-1/2}(\bbeta-\bbeta_0)'V_{12}'V_{11}^{-1}W_n-\frac{1}{2}n^{-1}W_n' V_{11}^{-1}\,W_n+o_P(n^{-1}),\]
with matrices
$V_{11}=(E\{g_j(X, Y, \bbeta)\,g_k(X, Y, \bbeta)'\})$, $V_{12}=-\frac{\partial}{\partial \beta_k}E\{g_j(X, Y, \bbeta)\}),$
and vector $W_n=n^{-1/2}\sum_{i=1}^n g(X_i, Y_i, \bbeta)).$

Then from $\log \mathfrak{R}(\bbeta)=n \Gamma_n(\bbeta)$ we have the posterior \[\pi(\bbeta|data)=\pi(\bbeta)\,\mathfrak{R}(\bbeta)
\propto \exp\{-\frac{1}{2}(\bbeta-\hat \bbeta)' I_n (\bbeta-\hat \bbeta)+O_p(1)\},\]
where $I_n=n V_{12}' V_{11}^{-1} V_{12}$  and $\hat \bbeta$ is the empirical likelihood estimate.

\section{Numerical experiments}
\label{section:Applications}

In this section we demonstrate our approach to Bayesian mode regression through  two simulated and one real examples. The real example is based on the Western Electric Workers (WECO) dataset and investigates how the worker's gender, pre-employment test result and education, can affect productivity.

\subsection{Simulation example 1}
\label{subsec: Simulations parametric}

We consider a simulated data from the  model
\begin{equation}
y_i=\beta_0+ \beta_1 x_i+\epsilon_i,
\end{equation}
where $x_i\sim N(0,1)$ and $i=1, ..., n$, with $n=50, 100, 200$. We set $\bbeta=(1,2)$ and
 consider the following three specifications for the model error $\epsilon$:

\begin{itemize}
	\item Case 1: the standard normal distribution, $\epsilon_i \sim N(0,1)$ - a symmetric error distribution;
	\item Case 2: a Fisher's Z distribution, $\epsilon_i \sim 1/2 log Z$ with $Z\sim F_{2,2}$  - a skewed error distribution;
	\item Case 3: a normal distribution with normally distributed outliers (contaminants) centered at twice the distance between the true mode and the $99^{th}$ percentile of the original normal distribution and accounting
for 20\% of the total data points, $\epsilon_i \sim 0.80N(0,\frac{1}{4})+0.20N(2.5,\frac{1}{4})$ \citep{hedges2003} - an asymmetric error distribution.
\end{itemize}

We fit parametric Bayesian mode regression (labeled PBMR) for all the cases above.
 Then for demonstration and comparison purposes  we fit empirical likelihood based Bayesian mode regression (labeled ELBMR) for case 2 and nonparametric Bayesian mode regression (labeled NBMR) for case 3.

 For PBMR and ELBMR, we chose independent improper uniform priors for all the components of $\bbeta$ and we simulated realizations from the posterior distributions by means of a single-component Metropolis-Hastings algorithm. Each of the parameters was updated using a random-walk Metropolis algorithm with a Gaussian proposal density centered at the current state of the chain. The variance of the proposal density was determined to provide an acceptance rate close to the optimal acceptance rate as defined in \citet{roberts2001}. Convergence was assessed using time series plots and the R package boa \citep{smith2007}. The estimates are posterior means using 10,000 iterations of the MCMC sampler (after 10,000 burn-in iterations).

The  estimates for NBMR were obtained by fitting a truncated Dirichlet Process (DP) mixture model, which leads to a computationally straightforward approximation and can be easily implemented in the freely available WinBUGS software. Two parallel chains of equal length with different initial values were run for the model. The results were based on 10,000 iterations which followed a burn-in period of 40,000 for each chain.

\begin{table}[!ht]
\caption{Simulation Example 1: True parameter values (T.V.) and their posterior means, standard deviations (S.D.) and 95\% credible intervals}
\label{table: Simulation results}
\begin{center}
\small\addtolength{\tabcolsep}{-7pt}
\begin{tabular} {|>{\centering}p{1cm}|c|cc|cc|cc|cc|cc|}
\hline
\multicolumn{2}{|c|}{} &\multicolumn{6}{c|}{PBMR} &\multicolumn{2}{|c|}{ELBMR} &\multicolumn{2}{c|}{NBMR}\\ \hline
\multicolumn{2}{|c|}{}&\multicolumn{2}{c|}{Normal} &\multicolumn{2}{c|}{Skewed} &\multicolumn{2}{c|}{Asymmetric}  &\multicolumn{2}{|c|}{Skewed} &\multicolumn{2}{c|} {Asymmetric} \\
\hline
n& &$\beta _{0}$ &$\beta _{1}$ &$\beta _{0}$ &$\beta _{1} $ &$\beta _{0}$ &$\beta _{1}$ &$\beta _{0}$ &$\beta _{1}$ &$\beta _{0}$ &$\beta _{1}$\\ \hline
 \multirow{5}{*}{50}
&T.V	&1	&2	&1	&2	&1	&2	&1	&2	&1	&2	\\
&Mean	&0.92	&2.00	&1.07	&2.01	&0.96	&2.02	&1.01	&2.00	&1.09	&1.94\\	
&S.D.	&0.78	&0.77	&0.78	&0.49	&0.34	&0.24	&0.01	&0.01	&0.24	&0.19\\	
&95\% HPD	&(-0.6,2.1)	&(0.5,3.3)	&(-0.3,2.6)	&(1.2,3.1)	&(0.4,1.7)	&(1.6,2.5)	&(0.99,1.02)	&(1.99,2.01)
&(0.7,1.5)	&(1.5,2.3)\\											
\hline
 \multirow{5}{*}{100}
&T.V	&1	&2	&1	&2	&1	&2	&1	&2	&1	&2\\	
&Mean	&1.01	&2.10	&0.95	&1.89	&1.06	&1.94	&1.01	&2.00	&1.06	&2.00	\\
&S.D.	&0.18	&0.25	&0.52	&0.37	&0.98	&0.76	&0.01	&2	&0.14	&0.12\\	
&95\% HPD	&(0.6,1.3)	&(1.6,2.6)	&(0.0,1.9)	&(1.2,2.6)	&(-0.7,2.9)	&(0.5,3.3)	&(0.99,1.02)	&(1.99,2.01)	 &(0.8,1.3)	 &(1.8,2.2)\\	
\hline
 \multirow{5}{*}{200}
&T.V	&1	&2	&1	&2	&1	&2	&1	&2	&1	&2	\\
&Mean	&1.26	&1.99	&1.00	&1.99	&1.06	&1.96	&1.01	&2.00	&1.04	&1.91	\\
&S.D.	&0.86	&0.52	&1.29	&0.75	&0.82	&0.42	&0.01	&0.01	&0.07	&0.06\\	
&95\% HPD	&(-0.5,2.8)	&(0.9,3.0)	&(-1.3,3.5)	&(0.6,3.3)	&(-0.4,2.6)	&(1.2,2.7)	&(0.99,1.02)	&(1.99,2.01)	 &(0.92,1.19)	 &(1.78,2.03)\\	
\hline
\end{tabular}
\end{center}
\end{table}

Table \ref{table: Simulation results} compares the posterior means with the true values of $\beta_0$ and $\beta_1$ and also gives standard deviations and 95\% credible intervals for each of the models considered in this example .

As expected, the PBRM works well as all the absolute biases for the estimated parameters turn out to be in the range [0.01, 0.26]. Furthermore, under both ELBMR and NBRM, the true values for both $\beta_0$ and $\beta_1$ are recovered successfully indicating that the methods also work well. However, it should be noted that the standard deviations for both parameters are smaller than in the PBMR, giving shorter confidence intervals.

\begin{figure}[!ht]
\caption{Plots showing the empirical samples from the joint distributions of mode  regression parameters}
\label{figure: joint posterior}
\centering \SetFigLayout{1}{3}
\subfigure[Symmetric error] {\includegraphics[height=5cm]{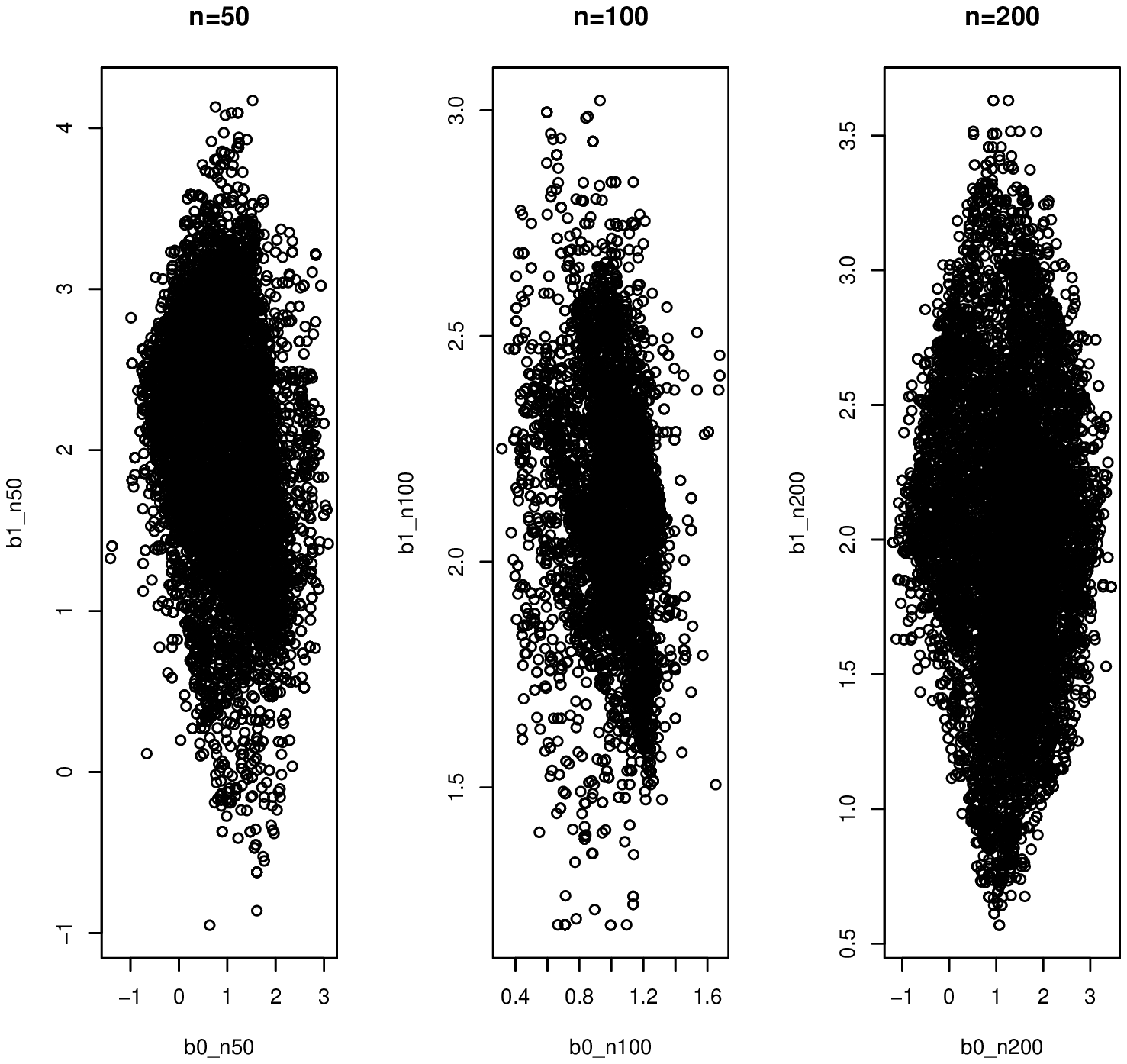}}\\
\subfigure [Skewed error]{\includegraphics[height=5cm]{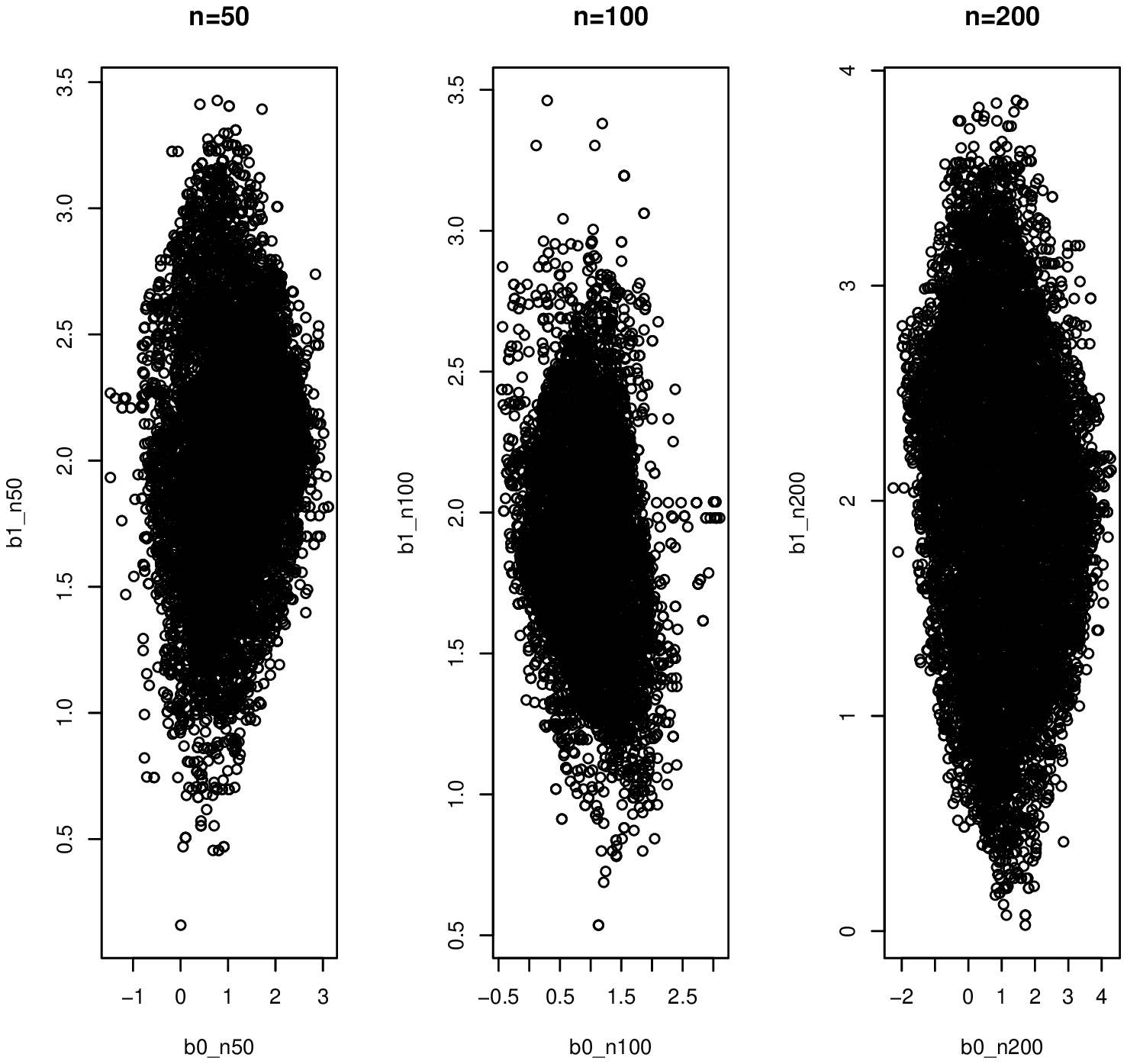}}\\
\subfigure[Asymmetric error]{\includegraphics[height=5cm]{joint_sym_sym}}
\end{figure}

Figure \ref{figure: joint posterior} exhibits the empirical samples from the joint posterior distributions of the PBMR parameters, which were obtained using the output of the MCMC sampler for the regression parameters $\widehat{\beta_0}$ and $\widehat{\beta_1}$. These samples can be used to obtain a consistent estimator of the covariance or correlation structure of the parameter estimators, which is difficult to estimate under the classical approach. For example in case (a), with sample size n=100,
\[ \widehat{Cov}{\hat{\beta_0} \choose \hat{\beta_1}}=
\begin{pmatrix}
  3 & -1\\
  -1 & 6 \\
\end{pmatrix}
\]

\newpage

\subsection{Simulation example 2}
\label{subsec: Simulations compare}

In this section we present the results of a second simulation example with the aim of comparing the performance of our approach with the classical mode regression approach. Specifically, we replicate the simulation study in \citeauthor*{kemp2012}(2012), but only for a sample of size $250$, and compare their results with the results obtained under our Bayesian mode regression approach.

Simulation data are generated by the simple linear model

\begin{equation}
y_i=\beta_0+ \beta_1 x_i+(1+vx_i)\epsilon_i,
\end{equation}
where $x_i$ are generated from a $\chi^2_{(3)}$ distribution, scaled to have variance 1, and $\epsilon_i$ are generated
as independent draws from a re-scaled log-gamma random variable,

\begin{equation}
\epsilon_i = -\lambda\, ln(Z_i)
\end{equation}
where $Z$ follows a gamma distribution with mean $1$ and scale parameter $\frac{1}{\alpha}$, to ensure that $\epsilon_i$ has zero mode. Furthermore, we set $\lambda=[(1+2E(x_i)v+E(x^2_i)v^2)\psi(\alpha)]$ \footnote{$\psi(\cdot)$ is the trigamma function} to ensure that the unconditional variance of
the error $(1+vx_i)$ is equal to one.

The study was performed for $\alpha\in \{0.05, 5\}$ and for $v\in \{0, 2\}$. Table \ref{table: Simulation results2} compares the 95\% credible intervals for the estimates obtained under  PBMR and NBMR  with the 95\% confidence intervals for the estimates under the two classical mode regression models: Mode 1.6 and Mode 0.8. Mode 1.6 and Mode 0.8 correspond to $k=1.6$ and $k=0.8$ respectively in the bandwidth selection rule, bandwidth=$k\,mad\,n^{-0.143}$,
with $mad$= the median of the absolute deviation from the median of ordinary least squares regression residuals.

\begin{table}[!ht]
\caption{Simulation Example 2: Comparison between Classical and Bayesian approach for mode regression}
\label{table: Simulation results2}
\begin{center}
\small\addtolength{\tabcolsep}{-5pt}
\begin{tabular} {|>{\centering}p{1cm}|>{\centering}p{1cm}|>{\centering}p{1cm}|c|c|c|c|c|}
\hline
\multicolumn{3}{|c|}{} &PBMR  & NBMR &Mode 1.6 &Mode 0.8 \\\hline
$\alpha$ &n & &95\% HPD  & 95\% HPD &95\% CI &95\% CI  \\\hline
\multirow{4}{*}{5.00} 	
&\multirow{2}{*}{0}
&$\beta _{0}$        &(-0.37,0.29)           &(-0.21,0.36)	        &(-0.31, 0.41)        &(-0.69, 0.75)	\\
&&$\beta _{1}$       &(0.82,1.28)	         &(0.89,1.32)	        &(0.77, 1.24)         &(0.56,1.45)	\\
\cline{2-7}
&\multirow{2}{*}{2}
&$\beta _{0}$        &(-0.06,0.07)	         &(-0.03,0.21)	        &(-0.15,0.23)         &(-0.25,0.29)	\\
&&$\beta _{1}$       &(0.99,1.14)            &(0.80,1.22)	        &(0.63,1.37)          &(0.48,1.53)	\\
 \hline
 \multirow{4}{*}{0.05} &\multirow{2}{*}{0}
&$\beta _{0}$        &(0.00, 0.14)	         &(-0.03,0.07)	        &(0.12,0.42)         &(-0.09,0.35)	\\
&&$\beta _{1}$       &(0.95,1.13)	         &(0.95,1.06)	        &(0.90,1.11)         &(0.87,1.17)	\\
\cline{2-7}
	&\multirow{2}{*}{2}
&$\beta _{0}$        &(0.02,0.08)	         &(0.04,0.09)	        &(0.09,0.29)         &(0.01,0.21)	\\
&&$\beta _{1}$       &(0.99,1.08)	         &(0.97,1.04)	        &(0.91,1.19)         &(0.85,1.19)	\\
\hline
\end{tabular}
\end{center}
\end{table}

The results of the analysis suggest that the Bayesian mode regression estimates are strong competitors of the classical mode regression estimates  since in almost all the examples both PBMR and NBMR estimators outperform the two classical estimators.

Finally, as also evident form \citeauthor*{kemp2012}(2012), the selection of the value/prior for $\sigma$ plays an important role on the precision of the parameters, an issue that is less restrictive under NBMR.

\subsection {Productivity of Western Electric Workers - WECO}
\label{subsection:weco parametric}

To illustrate the applicability  of our approach we consider a model for predicting the productivity of newly
hired Electric workers in a manufacturing firm. Productivity ($y_i$) was modeled as a function of a gender indicator ($sex_i$), the score on a physical dexterity exam administrated prior to employment ($dex_i$) and the years of education ($lex_i$).

\begin{equation}
y_i=\beta_0+\beta_1 sex_i+\beta_2 dex_i+\beta_3 lex_i+\beta_4 lex_i^{2}+\epsilon_i
\end{equation}

\begin{figure}
\centering
\caption{Density plot for WECO data}
\label{fig:Density plot for weco data}
\includegraphics[width=10cm]{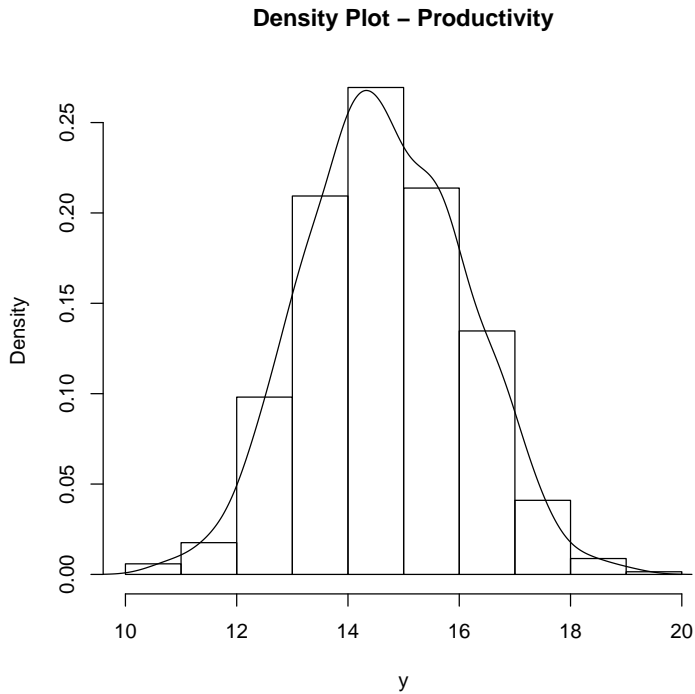}
\end{figure}

The data come originally from the study of \citet{klein1991}, but have been modified over the years to heighten the pedagogical impact. Figure \ref{fig:Density plot for weco data} presents the density plot for productivity which is unimodal and
almost symmetric (skewness =0.069).

While the productivity levels range from 10.5 to 19.1, one is interested in how the typical productivity level is affected by the model covariates. To estimate this effect we apply our PBMR model to estimate the model parameters, $\beta_0, \beta_1,\beta_2, \beta_3$ and $\beta_4$. The output was obtained by running the sampler for 50,000 cycles after a burn-in of 100,000, to ensure convergence and mixing. Table \ref{tab: Weco deta - results} summarizes the results.

The results indicate that on average the mode productivity level of a female worker, who scores zero on her physical dexterity exam and has zero years of education is 4.93 units. Furthermore, it can be concluded that on average the most frequent productivity level is lower for a male worker, while it is higher for workers with a higher exam score. Finally, it is deduced that an additional year of education contributes positively to the level of mode productivity.

Given that under the PBMR a relatively wide credible interval is obtained for the some model parameters we also fit a NBMR to the WECO dataset. Again, two parallel chains of equal length with different initial values were run for the model. The results were based on 20,000 iterations which followed a burn-in period of 50,000 iterations for each chain. As illustrated in  Table \ref{tab: Weco deta - results}, the results obtained under the NBMR are similar to the results obtained under PBRM, but now the confidence intervals are much smaller.

\begin{table}[!ht]
\caption{Model parameters and their estimated posterior means, standard deviations (S.D.) and 95\% Credible
intervals  for the WECO data}
\label{tab: Weco deta - results}
\begin{center}
\begin{tabular} {|c|c|c|c|c|c|c|}
\hline
&\multicolumn{3}{c|}{PBMR} &\multicolumn{3}{c|}{NBMR}\\
\hline
Parameter &Mean &S.D. &95\% HPD &Mean &S.D. &95\% HPD \\ \hline
$\beta _{0}$ & 4.93 &8.13 &(-12.6,19.8)& 4.10 &1.11 &(2.56, 6.52)\\
$\beta _{1}$ &-0.71 &0.46 &(-1.58,0.21) &-0.84 &0.08 &(-1.03,-0.73)\\
$\beta _{2}$ &0.12 &0.03 &(0.06,0.18) &0.12 &0.005 &(0.11,0.12) \\
$\beta _{3}$ &0.87 &1.27 &(-1.44,3.51) &1.08 &0.18 &(0.69,1.37) \\
$\beta _{4}$ &-0.04 &0.05 &(-0.14,0.06 )&-0.05 &0.008 &(-0.06,-0.03 )\\
\hline
\end{tabular}
\end{center}
\end{table}

\section{Conclusions}
\label{sec: Conclusions}

In this paper  we introduce a novel  Bayesian  mode regression framework which includes three approaches: a parametric  method, a nonparametric  method and an empirical likelihood based method, as in the area of mode regression, there is no literature from a Bayesian perspective.  We demonstrate that our estimates are consistent and asymptotically normal under rather standard conditions, even under misspecification of the likelihood function.   The approaches are  easy to implement and have proper inference tools as well as credible intervals available irrespective of the sample size. The numerical studies suggest that the proposed Bayesian mode regression estimates are strong
competitors of the classical mode regression estimates.

\appendix
\gdef\thesection{Appendix \Alph{section}}
\section{}
\label{appendix}

\textbf{Proof of theorem 1}

For any $\sigma>0$ and $m>p$, the moments of posterior distribution is given by
\[E[|\bbeta|^{\gamma}|\sigma, \bmy]=\int \prod_{j=0}^p |\beta_j|^{r_j} \frac{e^n}{(2 \sigma)^n}\,\sum_{i=1}^n
\exp[-I[|y_i-x_i'\bbeta|<\sigma]]
\,d \bbeta.\]
Noting that $\sum_{i=1}^n \exp[-I[|y_i-x_i'\bbeta|<\sigma]]$ is always a constant whether $|y_i-x_i'\bbeta|<\sigma$
 or not ($i=1, ..., n$).
Suppose that the coefficient matrix $X=(\bmx_1, \bmx_2, ...,\bmx_p)$ of mode regression equations
$y_i=\bmx_i^T \bbeta +\epsilon_i$ is a full rank matrix with rank $p$, then there is a subset of $p$
constrains  $|y_i-x_i'\bbeta|<\sigma$ ($i=1, ..., n$) to provide $0<|\beta_j|<B_j<\infty$ $(j=0, 1, ..., p-1)$,
even if some of $|y_i-x_i'\bbeta|<\sigma$ are true and some are not. Therefore, \[E[|\bbeta|^{\gamma}|\sigma,
\bmy]=const.\,\int_{-B_0}^{B_0}\int_{-B_1}^{B_1}...\int_{-B_p}^{B_p} \prod_{j=0}^p |\beta_j|^{r_j} \,d
\bbeta,\]
which is finite.

\bibliographystyle{Chicago}
\bibliography{mode}


\end{document}